\documentclass[12pt]{article}
\usepackage{a4}

\begin{document}

\begin{titlepage}

\begin{center}
{\Large\bf A Model for Neutrino Masses}
\vfill

{\bf C. Jarlskog}\\[0.3cm]

%
%
{Division of Mathematical Physics\\
LTH, Lund University\\
Box 118, S-22100 Lund, Sweden}

\end{center}
\vfill
\begin{abstract}

I present a model for neutrino masses based on a hypothesis proposed by
Friedberg and Lee.

\end{abstract}
\vfill
{Invited talk presented at the IV International Workshop on Neutrino Oscillations,
April 15-18 (2008) Venice, Italy}
\vfill
\end{titlepage}

\section{Prelude}

I am delighted to be giving a talk here in Italy 
on this very special day, April 15,
Leonardo da Vinci's birthday. Through my talk,
I will turn to him for guidance and understanding.

\section{Historical Perspective}

Let me first start with a bit of history. 
The year 1957 was a wonderful 
neutrino year due to the discovery of parity violation in weak interactions.
See the Nobel lectures by Lee and Yang
(\cite{lee57}, \cite{yang57}), both easily accessible on
the internet. The only weak interactions studied at that time
were decay processes where one or two 
neutrino particles were emitted. 
A remarkable feature of parity violation was
that it was compatible with being
maximal. A beautiful idea was then put forward that 
attributed this feature to an intrinsic property of 
the neutrino, its handedness. Consequently, the neutrino
was automatically massless.
This proposition came from no less distinguished scientists than
Lee and Yang \cite{leeyang}, Landau \cite{landau} and Salam
\cite{salam}, and perhaps even others. For example, Landau wrote
about what he called the longitudinally polarized, or more
simply longitudinal neutrino:

"In the sense of the usual scheme this would signify
that the neutrino is always polarized in the direction
of its motion (or in the opposite direction). The polarization
of the antineutrino is correspondingly reversed. According to
this model the neutrino is not a truly neutral particle".

He emphasized that "In the usual theory the neutrino mass
is zero, so to say, accidentally". 

Lee and Yang referred to it as the
two-component neutrino theory. This theory was largely
taken for granted for several decades and became 
a part of the 
weak-interactionists' scientific heritage. A lovely,
predictive theory that led to simplicity. That explains
why the founders of the Standard Electroweak Model chose
not to introduce right-handed neutrinos. The same goes
for the Grand Unified Georgi-Glashow Model. It would
have been a trivial task for the authors in question
to extend their models by adding in right-handed
neutrinos. One would have had what we could call
Slightly Non-minimal Standard Model \cite{cejasing}.
Nowadays, due to neutrino oscillations, 
we generally believe that neutrinos are
massive and therefore the two-component
neutrino theory is not valid. However, we do not understand
why and how the neutrinos are massive.

\vspace{.2cm}
Leonardo would tell us:
 
\vspace{.2cm}
"Although nature commences with reason and ends in experience 
it is necessary for us to do the opposite, 
that is to commence with experience and from this 
to proceed to investigate the reason."

\vspace{.2cm}
Thank you, Leonardo! We will keep your advice in mind.
Perhaps someday we will find the reason.

\section{The Friedberg-Lee Hypothesis}

Recently Friedberg and Lee have, in a series of articles
(\cite{friedlee06}, \cite{friedlee08}, \cite{friedlee07})
introduced and applied a new hypothesis as I will briefly
describe now.

These authors consider a model with three standard
families, each containing an 
up-type, a down-type as well as a 
charged lepton and a neutrino. There are thus
four "sectors", up-type, etc. For each sector,
Friedberg and Lee introduce Dirac mass terms of the kind
\begin{equation}
L_{mass} = \overline{\psi_j}M_{jk}\psi_k
\end{equation}
where, for the up-type quarks $\psi_j, j=1,2,3$ denote
three fields with quantum numbers of the up-type quarks
and $M$ is the corresponding mass matrix. The other
three sectors are treated similarly. Thus, all in all, there
are four such mass terms. Friedberg and Lee require each such
mass term to be invariant under a corresponding translation
of the fermion fields given by 
\begin{equation}
\psi_j \rightarrow \psi_j + \eta_j \theta
\label{flsym}
\end{equation}
Here, $\theta$ is a space-time independent four-component 
Grassmann number, $\theta_a\theta_b + \theta_b \theta_a =0$,
$a,b$ being Dirac indices. Furthermore $\eta_j$ are three
complex numbers. Note that there are four
$\theta$'s, one for each sector, and each sector has its
own set of $\eta$'s. 

Imposing the condition (\ref{flsym}) leads to the constraint

\begin{equation}
M_{jk} ~\eta_k =0
\end{equation}

If this condition were to be valid for
arbitrary $\eta$'s (that is to be a perfect symmetry) 
the mass matrix would vanish
identically. This is obviously too restrictive. Therefore,
Friedberg and Lee require that there exist a set of 
$\eta$'s for which the above equation is valid.
This implies that the
mass matrix must have a zero eigenvalue. 
This is good news because in each of the three
charged sectors there is one member that is much
lighter than the other two: 
the mass of the up-quark is much smaller than that 
of charm and top. Similarly, 
the electron is much lighter than the muon and 
the tau lepton. To a somewhat lesser extent the same
pattern also shows up in the down-type sector.
Similarly, in the framework of Friedberg and Lee,
the neutrinos are Dirac particles and one of them is
massless.

\vspace{.2cm}
Perhaps Leonardo would say:

\vspace{.2cm}
"Life is pretty simple: You do some stuff. Most fail. 
Some work. You do more of what works. 
If it works big, others quickly copy it."

\vspace{.2cm}
Thank you Leonardo! I think that the Friedberg-Lee idea
works big. It deserves to be copied. I'll wander down
their path hoping to find an exciting new vista.

\section{Choices within Slightly Non-minimal Standard Model}

The Standard Model is a chiral theory. Thus the fermion mass
terms are of the form
\begin{equation}
L^{SM}_{mass} = \overline{\psi_{jL}}M^{Dir}_{jk}\psi_{kR} + h.c.
\end{equation}
for the quarks and charged leptons. Here $Dir$
stands for Dirac. In Slightly Non-minimal Standard Model,
i.e., with right-handed neutrinos, we generally expect
to have, in addition to
Dirac mass terms, also Majorana mass 
terms for the right-handed neutrinos because there is no
symmetry that forbids such terms. 

Consider the fermionic translation in (\ref{flsym}). In a chiral
theory, there are several possibilities for introducing it
because the left-handed and the right-handed fermions are
independent of each other. Therefore, one may choose to
introduce the Friedberg-Lee condition for:
\vspace{.5cm}

1. only the left-handed fermions

2. only the right-handed fermions

3. both chiralities.
\vspace{.5cm}

One could also choose some combinations of the above
alternatives. Let us consider not only the mass terms
but the entire Lagrangian of this Slightly Non-minimal Standard
Model. Option 1, applied to the entire Lagrangian,
would kill all interactions of the corresponding 
fermions with the $W$-bosons.
In option 2, the interactions with $W$-bosons are 
not affected but 
the interactions of right-handed fermions with
the $B$-boson will be forbidden. This is a less severe
restriction than we had in the former case. 
The right-handed neutrino
sector is very special because right-handed
neutrinos don't interact
with the gauge bosons. Furthermore, the kinetic terms 
of fermions change, under the above translation, 
by a total derivative whereby
the resulting action is invariant.
This means that all the terms in the Lagrangian, except the
interactions with the Higgses and Majorana
mass terms, are automatically invariant under the
Friedberg-Lee translation of the right-handed
neutrino fields. In other words, the right-handed neutrinos
constitute a basis for maximal symmetry. This was the
motivation for the work in \cite{ceja08} that I would like
to discuss now.

\vspace{.2cm}
Leonardo, what would you say about this? I hear you say

\vspace{.2cm}
"Human subtlety will never devise an invention more beautiful, 
more simple or more direct than does nature because in her 
inventions nothing is lacking, and nothing is superfluous."

\vspace{.2cm}
Yes, Leonardo. You are absolutely right. Please forgive me
for pursuing an idea that is by no means more beautiful
or more direct than does nature. It may be wrong but at least
it is simple. 

\section{The $\nu_R$-sector}

Following the discussion above, we assume that the
Lagrangian is invariant under the translation
\begin{equation}
\nu_{Rj} \rightarrow  \nu_{Rj} + \eta_j \theta
\label{nu-rsym}
\end{equation}
Again $\eta_j, ~j=1-3$ are complex numbers, and
$\theta$ is a space-time independent Grassmann number.
As noted before, this translation is a symmetry of the Standard 
Model excluding the interactions of the neutrinos with
the Higgs fields (which generate the Dirac mass terms) 
and the Majorana mass terms.
Therefore, requiring  
the above translations to be a symmetry (i.e., to be valid
for arbitrary $\eta$) of the entire action  
immediately would yield that the neutrino masses
are all zero. This result is interesting because it could perhaps
be the reason why the neutrino masses are small - 
they start off by being zero in the symmetry limit! One could
imagine that there is a scenario beyond the Standard Model in
which the above symmetry holds and somehow in the broken
version the $\eta$'s get fixed. 

The neutrino mass matrix, including Majorana terms, is given
by
\begin{equation}
L^{(\nu)}_{mass}= -{1 \over 2} (\overline{\nu^{}_L}, \overline {\nu^C_R}) 
{\cal M} \left( \begin{array}{c}
\nu_L^C \\ \nu_R \end{array} \right) + h.c.
\end{equation}
where
\begin{equation} \nu_X = \left(
\begin{array}{c} \nu_{1X} \\ \nu_{2X} \\ \nu_{3X}
\end{array} \right) 
\end{equation}
$X=L, R$. Here
${\cal M}$ is the six-by-six neutrino mass matrix
given by
\begin{equation} 
{\cal M}  = \left(
\begin{array}{cc}
0 & A  \\
A^T & M 
\end{array}
\right)
\end{equation}
${\cal M}$ is a 
symmetric matrix; 
$A$ and $M$ being respectively the three-by-three 
Dirac and Majorana mass matrices and $T$ stands for transpose. 

Requiring that the translation
$\nu_{Rj} \rightarrow  \nu_{Rj} + \eta_j \theta$ 
leave the Lagrangian invariant yields
\begin{equation}
A_{jk} \eta_k =0, ~~~
M_{jk} \eta_k =0
\end{equation}
We may now redefine the right-handed neutrino fields
by
\begin {equation}
\nu_R \rightarrow U \nu_R
\end{equation}
where $U$ is a three-by-three unitary matrix.
Under this transformation the Lagrangian retains its
form. All that happens is a redefinition of the mass matrices 
\begin{equation}
M \rightarrow M^\prime= U^{\star} M U^\dagger, ~~~
A \rightarrow A^\prime= A U^\dagger
\end{equation}
The Friedberg-Lee condition now reads
\begin{equation}
A^\prime_{jk} \eta^\prime_k =0, ~~~
M^\prime_{jk} \eta^\prime_k =0
\end{equation}
where $\eta^\prime = U \eta$. 
We may choose $U$ such that $M^\prime$ is diagonal. 
$M^\prime$ has a zero
eigenvalue. Therefore, by permutation of the right-handed neutrino
fields (which leaves the rest of Lagrangian invariant) we
may write $M^\prime$ in the form  
\begin{equation}
M^\prime = \left( \begin{array}{ccc}
M_1 &0 & 0\\ 
0 & M_2 & 0 \\
0 & 0 & 0
\end{array} \right) 
\end{equation}
Assuming that $M_1$ and $M_2$ are nonzero, we find
that the vector $\eta^\prime$ is proportional to $(0, 0, 1)$.
This in turn implies that the third column in matrix 
$A^\prime$ is zero. Therefore it is of the form
\begin{equation}
A^\prime = \left( \begin{array}{ccc}
a_{11} &a_{12} & 0\\ 
a_{21} & a_{22} & 0 \\
a_{31} & a_{32} & 0
\end{array}
\right)
\end{equation}
Thus a remarkable consequence of the Friedberg-Lee condition
is that one of the three right-handed neutrinos simply
decouples, in the sense that it has only gravitational
interactions due to its kinetic energy term.
We end up with three left-handed neutrinos
and just two right-handed neutrinos. The mass matrix is then of
the form
\begin{equation}
{\cal M} = \left( \begin{array}{ccccc}
0 & 0 & 0 & a_{11} &a_{12} \\ 
0 & 0 & 0 & a_{21} & a_{22}  \\
0 & 0 & 0 & a_{31} & a_{32} \\
a_{11} & a_{21} & a_{31} & M_1 & 0 \\
a_{12} & a_{22} & a_{32} & 0  & M_2
\end{array}
\right)
\end{equation}
To sum up, in this model there are two massless
neutrinos one of them being a 
non-interacting massless right-handed neutrino and
the other one
a massless interacting neutrino. 
A natural scenario 
would be to have small values for
magnitudes of $M_1$ and $M_2$ because in the symmetry limit
these are zero. In the limit $M_1, M_2 \rightarrow 0$
one obtains three Dirac neutrinos, one of them being massless.
The model may or may not violate CP symmetry depending
on how the charged lepton mass matrix looks like.

However, since we don't really know what
is natural or not, we should also keep in mind that 
the see-saw mechanism is also allowed. It is obtained 
by taking the magnitudes of $M_1$ and $M_2$ to be very
large. Then there will be two very heavy neutrinos
and three light ones, one of them being 
massless. The phenomenology of  
see-saw models with two right-handed neutrinos
has been the subject of several studies in the literature
(see, for example \cite{ross} and \cite{ibarra}
and references cited therein). See also Micha Shaposhnikov's 
contribution to this meeting \cite{micha}
 
\vspace{.2cm}
I hear Leonardo declaring.

\vspace{.2cm}
"The noblest pleasure is the joy of understanding."

\section{Lorentz invariance} 
 
I would like to mention that fermionic translations
are far more subtle than their
bosonic counterparts, such as 
$\phi \rightarrow \phi + v$, or $A_\mu \rightarrow
A_\mu + \partial_\mu \Lambda$. For example, under 
Lorentz transformations
$\theta$ must transform as a fermionic field, i.e., the
transformation $\nu_R \rightarrow S ~\nu_R$ 
implies $\theta \rightarrow S
~ \theta$, where $S$ is the appropriate transformation
matrix.  
A nonzero vacuum expectation value of a fermionic operator
will break Lorentz invariance. One could imagine
that the Friedberg-Lee translation is actually driven
by the requirement of Lorentz invariance, such as
in a brane world picture, where Lorentz invariance is
violated in the bulk but restored on the brane.

\vspace{.2cm}
I am not sure that Leonardo would agree with this.
Perhaps he would say:

\vspace{.2cm}
"Blinding ignorance does mislead us. 
O! Wretched mortals, open your eyes!"

\section{Acknowledgements}

My most sincere thanks go to Milla Baldo-Ceolin, for
inviting me to this meeting and asking me to speak.
For me she is a role model, by her kindness and dedication
to our field. I would like also to thank you, 
Leonardo, for your fantastic
multidisciplinary contributions
and for being born on 15 April. It has been great to
get to know you better.


\begin{thebibliography}{99}
\baselineskip=15pt



\bibitem{lee57} T. D. Lee, Nobel Lecture (1957) 

\bibitem{yang57} C. N. Yang, Nobel Lecture (1957)

\bibitem{leeyang} T. D. Lee and C. N. Yang,
Phys. Rev. 105 (1957) 1671

\bibitem{landau} L. Landau, Nucl. Phys., 3 (1957) 127

\bibitem{salam} A. Salam, Nuovo Cimento, 5 (1957) 299

\bibitem{cejasing} C. Jarlskog, Proceedings of the 25th International
Conference on High Energy Physics (Singapore, 1990), p. 61
(Eds. K. K. Phua and Y. Yamaguchi)

\bibitem{friedlee06} R. Friedberg and T. D. Lee, Chinese Phys. C
Vol. 30 (2007) 591; hep-ph/0606071

\bibitem{friedlee08} R. Friedberg and T. D. Lee, Annals of Phys.
Vol. 323 (2008) 1087; arXiv:0705.4156

\bibitem{friedlee07} R. Friedberg and T. D. Lee, arXiv:0709.1526

\bibitem{ceja08} C. Jarlskog, Phys. Rev. D77 (2008) 073002;
arXiv:0712.0903


 
\bibitem{ross} A. Ibarra and G.G. Ross, Phys. Lett. B591 (2004) 285

\bibitem{ibarra} A. Ibarra, JHEP 0601 (2006) 064

\bibitem{micha} M. Shaposhnikov, These Proceedings

 

\end{thebibliography}
\end{document}